\documentclass{article}

\usepackage{spconf,amsmath,graphicx}
\usepackage{epstopdf}

\usepackage{graphicx}
\usepackage{color}
\usepackage{placeins}
\usepackage{float}
\usepackage{tabularx,colortbl}
\graphicspath{{figures/Appendix/}}
\usepackage{amssymb} 
\usepackage{amsmath} 
\usepackage[breaklinks=true]{hyperref}
\usepackage{setspace}
\usepackage{cite}
\usepackage{algorithmic}
\usepackage{algorithm}
\usepackage{subfig}
\usepackage{epstopdf}
\usepackage{morefloats}
\usepackage[square, comma, sort&compress, numbers]{natbib}

\def \Cfi {C_{f}}
\def \Csi {C_{s}}

\DeclareGraphicsExtensions{.pdf,.png,.jpg}

\title{Deconvolution Using Projections Onto The Epigraph Set of a Convex Cost Function}
\name{Mohammad Tofighi, Alican Bozkurt, and A. Enis Cetin}
\address{Department of Electrical and Electronic Engineering, Bilkent University,  Ankara, Turkey\\
tofighi@ee.bilkent.edu.tr, alican@ee.bilkent.edu.tr, cetin@bilkent.edu.tr\\}
\begin{document}
%
\maketitle
\begin{abstract}
A new deconvolution algorithm based on orthogonal projections onto the epigraph set of a convex cost function is presented. In this algorithm, the dimension of the minimization problem is lifted by one and  sets  corresponding to the cost function are defined. As the utilized cost function is a convex function in $\mathbb{R}^N$, the corresponding epigraph set is also a convex set in $\mathbb{R}^{N+1}$. The deconvolution algorithm starts with an arbitrary initial estimate in $\mathbb{R}^{N+1}$. At each step of the iterative algorithm, first deconvolution projections are performed onto the epigraphs, later an orthogonal projection is performed onto one of the constraint sets associated with the cost function in a sequential manner. The method provides globally optimal solutions for total-variation, $\ell_1$, $\ell_2$, and entropic cost functions.
\end{abstract}
\section{Introduction}
\label{sec:Introduction}
A new deconvolution algorithm based on orthogonal Projections onto the Epigraph Set of a Convex cost function (PESC) is introduced. In Bregman's standard POCS approach \cite{Bregman,You82}, the algorithm converges to the intersection of convex constraint sets.  In this article, it is shown that it is possible to use a convex cost function in a POCS based framework using the epigraph set and the new framework is used in deconvolution.

Bregman also developed iterative methods based on the so-called Bregman distance to solve convex optimization problems \cite{Bre67}. In Bregman's approach, it is necessary to perform a Bregman projection at each step of the algorithm, which may not be easy to compute the Bregman distance in general \cite{Yin08,Gunay}.

In standard POCS approach, the goal is simply to find a vector, which is in the intersection of convex constraint sets \cite{GlobalSIP2013,Yamada,Kose2013,You82,Cen12,Sla08,Cetin89,Kose11,Cen81,Sla09,The11,censor1987optimization,Tru85,Com04,Com93,yamada2011minimizing,censor1987some,Sez82,censor1992proximal,Tuy81,censor1981row,censor1991optimization,Ros13}. In each step of the iterative algorithm an orthogonal projection is performed onto one of the convex sets. Bregman showed that successive orthogonal projections converge to a vector, which is in the intersection of all the convex sets. If the sets do not intersect iterates oscillate between members of the sets \cite{Gub67,Cet97}. Since, there is no need to compute the Bregman distance in standard POCS, it found applications in many practical problems. In this article, orthogonal projections onto the epigraph set of a convex cost functions are used to solve convex optimization problems instead of the Bregman distance approach.

In PESC approach, the dimension of the signal reconstruction or restoration problem is lifted by one and  sets corresponding to a given convex cost function are defined. This approach is graphically illustrated in Fig.\ref{app:convex}. If the cost function is a convex function in $\mathbb{R}^N$, the corresponding epigraph set is also a convex set in $\mathbb{R}^{N+1}$. As a result, the convex minimization problem is reduced to finding the $[\mathbf{w}^*,f(\mathbf{w}^*)]$ vector of the epigraph set corresponding to the cost function as shown in Fig. \ref{app:convex}. As in standard POCS approach, the new iterative optimization method starts with an arbitrary initial estimate in $\mathbb{R}^{N+1}$ and an orthogonal projection is performed onto one of the constraint sets. The resulting vector is then projected onto the epigraph set. This process is continued in a sequential manner at each step of the optimization problem. This method provides globally optimal solutions for convex cost functions such as total-variation \cite{Chambolle}, filtered variation \cite{Kos12}, $\ell_1$ \cite{Bar07}, and entropic function \cite{Kose2013}. The iteration process is shown in Fig. \ref{app:convex}. Regardless of the initial value $\underline{\mathbf{w}}_0$, iterates converge to $[\mathbf{w}^*,f(\mathbf{w}^*)]$ pair as shown in Fig. \ref{app:convex}.

The article is organized as follows. In Section \ref{sec:Convex Minimization}, the epigraph of a convex cost function is defined and the convex minimization method based on the PESC approach is introduced. In Section \ref{sec:Denoising using PESC}, the new deconvolution method is presented. The new approach does not require a regularization parameter as in other TV based methods \cite{Yamada,Chambolle,Com04}. In Section~\ref{sec:Simulation Results}, the simulation results and some deconvolution examples, are presented.

\vspace{-0.1cm}
\section{Epigraph of a Convex Cost Function}
\begin{figure}[ht!]
\begin{center}
\noindent
\includegraphics[width=60mm]{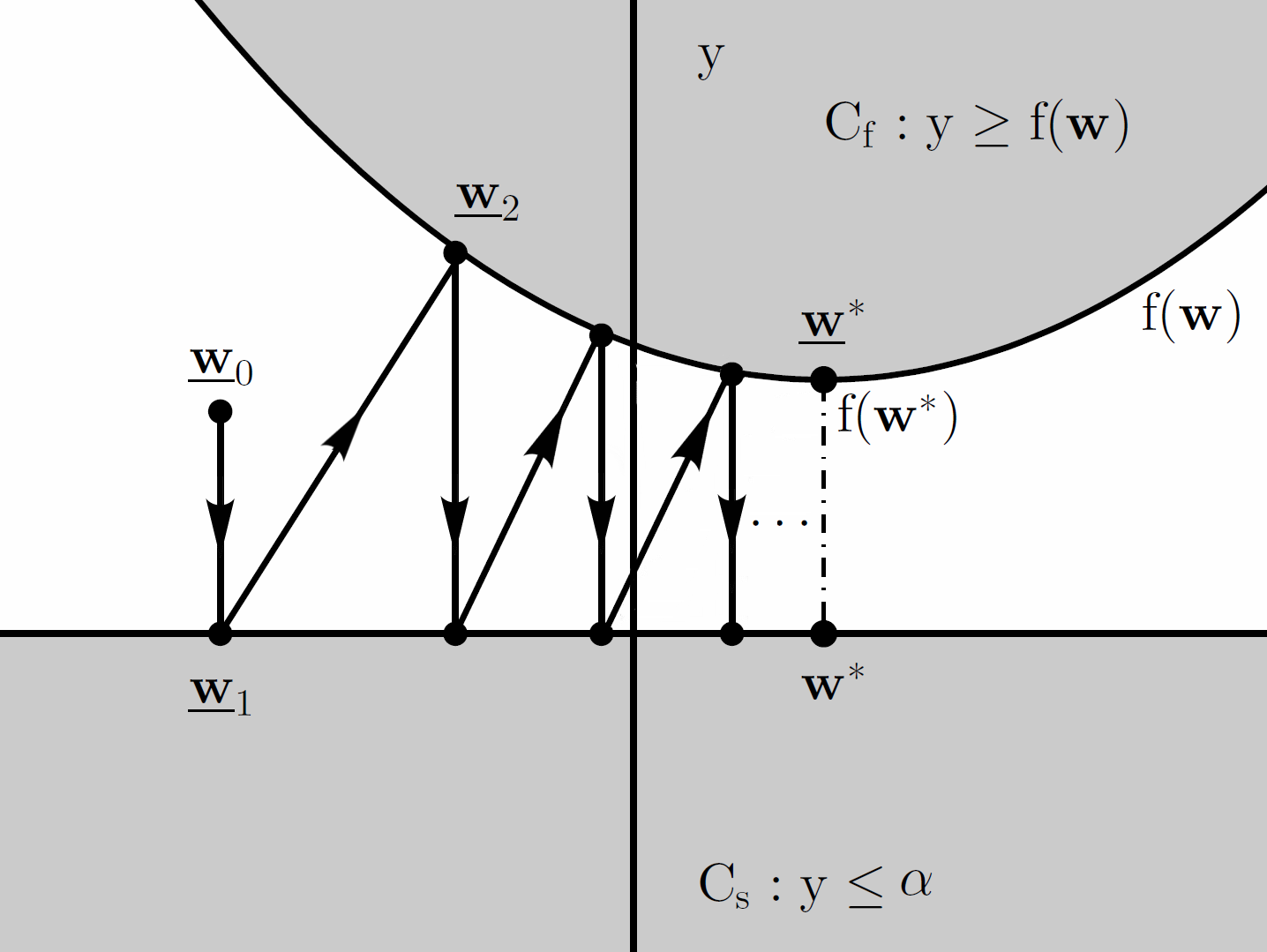}
\caption[Two projecting convex sets.]{Two convex sets $\Cfi$ and $\Csi$ corresponding to the convex cost function $f$. We sequentially project an initial vector $\underline{\mathbf{w}}_0$ onto $\Csi$ and $\Cfi$ to find the global minimum, which is located at $\underline{\mathbf{w}}^* = [{\mathbf{w}}^*~f(\mathbf{w^*})]^T$.}
\label{app:convex}
\end{center}
\end{figure}

\vspace{-0.1cm}
\label{sec:Convex Minimization}
Let us consider a convex minimization problem
\begin{equation}
\label{app:eq:c5}
\underset{\mathbf{w}\in \mathbb{R}^N}{\text{min}} f(\mathbf{w}),
\end{equation}
where $f:\mathbb{R}^N \rightarrow \mathbb{R}$ is a convex function.
We increase the dimension of the problem by one to define the epigraph set in $\mathbb{R}^{N+1}$ corresponding to the cost function $f(\mathbf{w})$ as follows:
\begin{equation}
\label{app:eq:c6}
\text{C}_f = \{\underline{\mathbf{w}} = [\mathbf{w}^T~y]^T : \mathrm{~} y\geq f(\mathbf{w})\},
\end{equation}
which is the set of $N+1$ dimensional vectors, whose $(N+1)^{st}$ component $y$ is greater than $f(\mathbf{w})$. We use bold face letters for $N$ dimensional vectors and underlined bold face letters for $N+1$ dimensional vectors, respectively. The second set that is related with the cost function $f(\mathbf{w})$ is the level set:
\begin{equation}
\label{app:eq:c7}
\text{C}_s = \{\underline{\mathbf{w}} = [\mathbf{w}^T~y]^T : ~ y\leq \alpha , ~\underline{\mathbf{w}} \in \mathbb{R}^{N+1}\},
\end{equation}
where $\alpha$ is a real number. Here it is assumed that $f(\mathbf{w})\geq \alpha$ for all $f(\mathbf{w})\in \mathbb{R}$ such that the sets $\Cfi$ and $\Csi$ do not intersect or the intersection contains a single vector. They are both closed and convex sets in  $\mathbb{R}^{N+1}$. Sets $\Cfi$ and $\Csi$ are graphically illustrated in Fig. \ref{app:convex}.

An important component of the PESC approach is to perform an orthogonal projection onto the epigraph set. Let $\underline{\mathbf{w}}_1$ be an arbitrary vector in $\mathbb{R}^{N+1}$. The projection $\underline{\mathbf{w}}_2$ is determined by minimizing the distance between $\underline{\mathbf{w}}_1$ and $\Cfi$, i.e.,
\begin{equation}
\label{app:eq:convex}
\underline{\mathbf{w}}_2 = \text{arg} \underset{\underline{\mathbf{w}}\in \text{C}_{\mathrm{f}}}{\text{min}} \|\underline{\mathbf{w}}_1 - \underline{\mathbf{w}}\|^{2}.
\end{equation}
Equation \ref{app:eq:convex} is the ordinary orthogonal projection operation onto the set $\mathrm{C_f} \in \mathbb{R}^{N+1}$.
In order to solve the problem in Eq. (\ref{app:eq:convex}) we do not need to compute the Bregman's so-called D-projection or Bregman projection. Projection onto the set $C_s$ is trivial. We simply force the last component of the $N+1$ dimensional vector to zero. In the PESC algorithm, iterates eventually oscillate between the two nearest vectors of the sets $\Csi$ and $\Cfi$ as shown in Fig. \ref{app:convex}. As a result, we obtain
\begin{equation}
\label{app:eq:convex2}
\underset{n \rightarrow \infty}{\text{lim}} \underline{\mathbf{w}}_{2n} = [\mathbf{w}^*~f(\mathbf{w}^*)]^T,
\end{equation}
where $\mathbf{w}^*$ is the N dimensional vector minimizing $f(\mathbf{w})$. The proof of Eq. (\ref{app:eq:convex2}) follows from Bregman's POCS theorem \cite{Bregman}. It was generalized to non-intersection case by Gubin et. al \cite{Gub67}. Since the two closed and convex sets $\Csi$ and $\Cfi$ are closest to each other at the optimal solution case, iterations oscillate between the vectors $[\mathbf{w}^*~f(\mathbf{w}^*)]^T$ and $[\mathbf{w}^*~0]^T$ in $\mathbb{R}^{N+1}$ as $n$ tends to infinity. It is possible to increase the speed of convergence by non-orthogonal projections \cite{Com93}.

If the cost function $f$ is not convex and have more than one local minimum then the corresponding set $\Cfi$ is not convex in $\mathbb{R}^{N+1}$. In this case iterates may converge to one of the local minima.

In current TV based deconvolution methods \cite{Com04, Com11, Com123}, the following cost function is used:
\begin{equation}
\label{app:eq:cost}
{{\text{min}} \|\mathbf{v} - \mathbf{w}\|}^2 + \lambda \text{TV}(\textbf{w}),
\end{equation}
where \textbf{v} is the observed signal. The solution of this problem can be obtained using the method in an iterative manner, by performing successive orthogonal projections onto $\Cfi$ and $\Csi$ , as discussed above. In this case the cost function is $f(\textbf{w}) = {\|\mathbf{v} - \mathbf{w}\|}^2_2 + \lambda \text{TV}(\textbf{w})$. Therefore,
\begin{equation}
\label{app:eq:8}
\Cfi = \{{\|\mathbf{v} - \mathbf{w}\|}^2 + \lambda \text{TV}(\textbf{w})\leq y\}.
\end{equation}
The deconvolution solutions that we obtained are very similar to the ones found by Combettes in \cite{Com123, Pesquet12} as both methods use the same cost function. One problem in TV based cost function is the estimation of the regularization parameter $\lambda$. One has to determine the $\lambda$ in an ad-hoc manner or by visual inspection. In the next section, a new deconvolution method with a different TV based cost function is described. The new method does not require a regularization parameter.
\vspace{-0.1cm}

\section{Deconvolution Using PESC}
\begin{figure}[ht!]
\begin{center}
\noindent
\includegraphics[width=70mm]{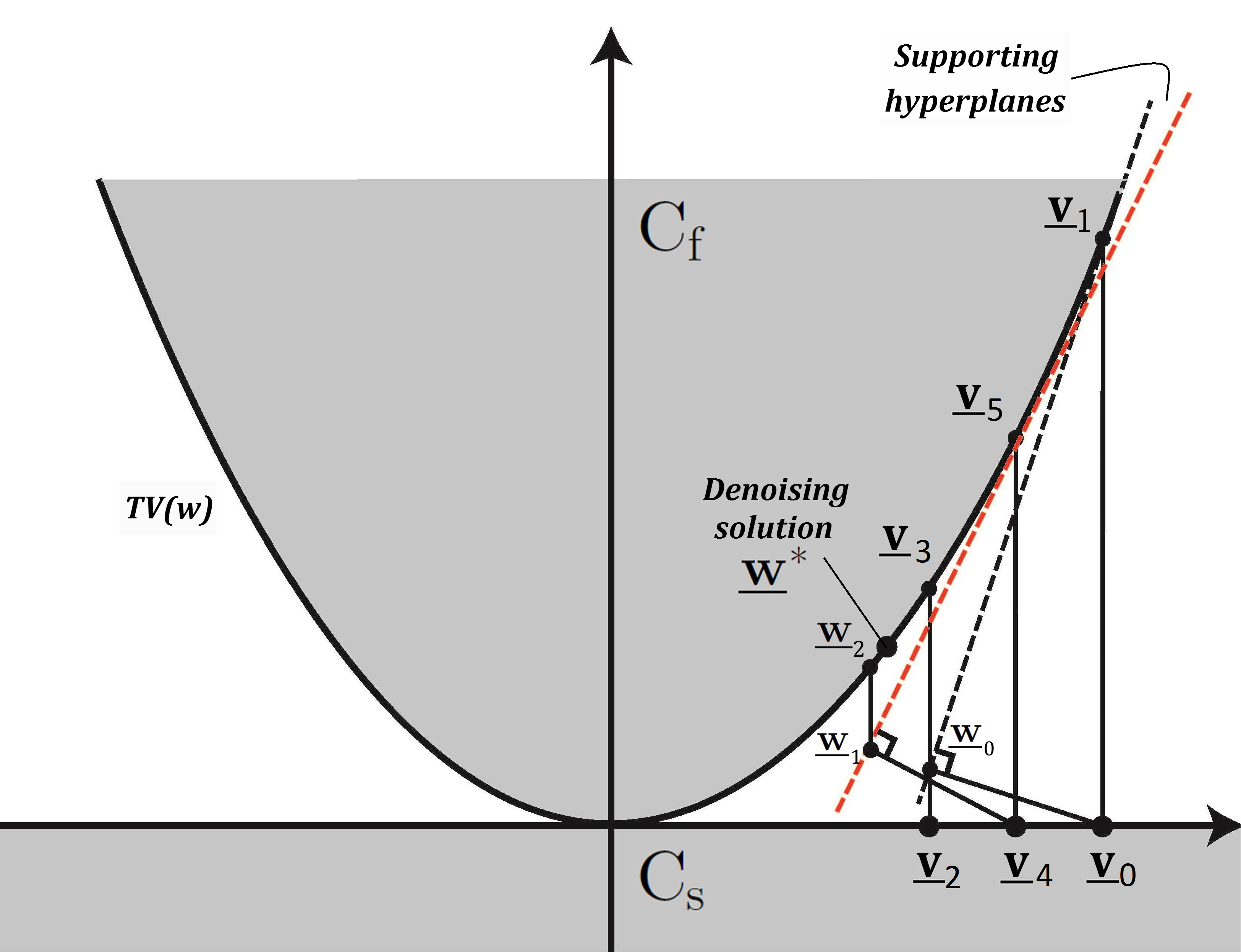}
\caption{Graphical representation of the minimization of Eq. (\ref{app:eq:6}), using projections onto the supporting hyperplanes of $\Cfi$. In this problem the sets $\Csi$ and $\Cfi$ intersect because $TV(\textbf{w})=0$ for $\textbf{w}=[0, 0,...,0]^{T}$ or for a constant vector.}
\label{app:convex1}
\end{center}
\end{figure}

\vspace{-0.1cm}
\label{sec:Denoising using PESC}
In this section, we present a new deconvolution method, based on the epigraph set of the convex cost function. It is possible to use TV, FV and $\ell_1$ norm as the convex cost function. Let the original signal or image be $\textbf{w}_{orig}$ and its blurred and noisy version be \textbf{z}:
\begin{equation}
\textbf{z} = \textbf{w}_{orig}\ast \textbf{h} + \boldsymbol{\eta},
\end{equation}
where $\textbf{h}$ is the blurring matrix and $\boldsymbol{\eta}$ is the additive Gaussian noise. In this approach we solve the following problems:
\begin{equation}
\label{app:eq:6}
\underline{\mathbf{w}}^{\star} = \text{arg} \underset{\underline{\mathbf{w}}\in \text{C}_{\mathrm{f}}}{\text{min}} \|\underline{\mathbf{v}} - \underline{\mathbf{w}}\|^{2},
\end{equation}
where, $\underline{\mathbf{v}}$ = [$\textbf{v}^{T}$\ 0] and $\Cfi$ is the epigraph set of TV or FV in $\mathbb{R}^{N+1}$.
The TV function, which we used for an $M\times M$ discrete image $\textbf{w} = [w^{i,j}],~~0\leq i,j\leq M-1~\in \mathbb{R}^{M\times M}$ is as follows:
\begin{equation}
\label{app:eq:7}
TV(\textbf{w}) = \sum_{i,j} (|w^{i+1, j} - w^{i, j}| + |w^{i, j+1} - w^{i, j}|).
\end{equation}
To estimate this problem we use POCS framework using the following sets:
\begin{equation}
\label{app:eq:c7}
\text{C}_i = \{\mathbf{w}\in \mathbb{R}^{N+1} | z_{i} = (\mathbf{w}\ast \mathbf{h})[i]\} \qquad i = 1, 2, ..., L,
\end{equation}
where L is the number of pixels and $z_{i}$ is the $i^{th}$ observation; and the epigraph set:
\begin{equation}
\label{app:eq:c6}
\text{C}_f = \{\underline{\mathbf{w}} \in \mathbb{R}^{N+1} | \underline{\mathbf{w}} = [\mathbf{w}^T~y]^T : \mathrm{~} y\geq TV(\mathbf{w})\}.
\end{equation}
Notice that the sets $\text{C}_i$ are in $\mathbb{R}^{N}$ and $\text{C}_f$ is in $\mathbb{R}^{N+1}$. However, it is straightforward to extend $\text{C}_i$'s to $\mathbb{R}^{N+1}$ and they are still closed and convex sets in $\mathbb{R}^{N+1}$.
Let us describe the projection operation onto the set $\Cfi = \{TV(\textbf{w})\leq y\}$. Notice that, this $\Cfi$ is different from Eq. (\ref{app:eq:8}). This means that we select the nearest vector $\underline{\mathbf{w}}^{\star}$ on the set $\Cfi$ to \textbf{v}. This is graphically illustrated in Fig. \ref{app:convex1}.
During this orthogonal projection operations, we do not require any parameter adjustment as in \cite{Chambolle}. The POCS algorithm consists of cyclical projections onto the sets $\text{C}_i$ and $\text{C}_f$.

Projection onto the sets are very easy to compute because they are hyperplanes:
\begin{equation}
\label{app:eq:9}
\textbf{v}_{r+1} = \textbf{v}_{r} + \frac{z_{i} - (\textbf{v}_{r}\ast \textbf{h})[i]}{\|\textbf{h}\|^2}\textbf{h}^{T},
\end{equation}
where $v_{r}$ is the $r^th$ iterate, $v_{r+1}$ is the projection vector onto the hyperplane $\text{C}_i$.
The pseudo-code of the algorithm is described in Algorithm \ref{fig:algo_1}.
\begin{algorithm}[H]
\begin{algorithmic}
\STATE{\textbf{Begin} $\textbf{z} \in \mathbb{R}^{N\times N}$, $\textbf{h} \in \mathbb{R}^{N_h\times N_h}$, $K \in \mathbb{Z^+}$ }
\STATE{$v \leftarrow z$}
\FOR{$k$ = 1 to K}
\FOR{$x$ = 1 to N}
\FOR{$y$ = 1 to N}
\STATE{$\textbf{v}(x-\lfloor N_h/2\rfloor ~to~x+\lfloor N_h/2\rfloor,y-\lfloor N_h/2\rfloor ~to~y+\lfloor N_h/2\rfloor) \leftarrow \textbf{v}(x-\lfloor N_h/2\rfloor ~to~x+\lfloor N_h/2\rfloor,y-\lfloor N_h/2\rfloor ~to~y+\lfloor N_h/2\rfloor)+ \frac{z(x,y)-v\ast h|_{x,y}}{\|h\|^2}h$}
\ENDFOR
\ENDFOR

\WHILE{$||\textbf{w}-\textbf{v}||> \epsilon$} 
\STATE{$\textbf{w} \leftarrow$ Project $\textbf{v}$ onto $C_f$}
\STATE{$\textbf{w} \leftarrow$ Project $\textbf{w}$ onto $C_s$}
\ENDWHILE
\ENDFOR
\end{algorithmic}
\caption{The pseudo-code for the deconvolution with PESC based algorithm}
\label{fig:algo_1}
\end{algorithm}
The sets $\text{C}_i$ and $\text{C}_f$ may or may not intersect in $\mathbb{R}^{N+1}$. If they intersect, iterates converge to to a solution in the intersection set. It is also possible to use hyperslabs instead of $C_{i, h} = \{\textbf{w}| z_i - \epsilon_i \leq (\textbf{w} \ast \textbf{h})[i]\leq z_i + \epsilon_i\}$ hyperplanes $\text{C}_i$ in this algorithm. In this case it is more likely that the closed and convex sets of the proposed framework intersect.

\emph{Implementation}: The sub-gradient projections of $v_{n}$ are performed as in Eq. \ref{app:eq:9}. Then after a loop of these projections are terminated, the PESC algorithm will be applied to the output $v_{n}$. The projection operation described in Eq. (\ref{app:eq:6}) can not be obtained in one step when the cost function is TV. The solution is determined by performing successive orthogonal projections onto supporting hyperplanes of the epigraph set $\Cfi$. In the first step, TV($\mathbf{v}_{0}$) and the surface normal at $\underline{\mathbf{v}}_{1}$ = [$\mathbf{v}_{0}^{T}$ \ TV($\mathbf{v}_{0}$)] in $\mathbb{R}^{N+1}$ are calculated. In this way, the equation of the supporting hyperplane at $\underline{\mathbf{v}}_{1}$ is obtained.
The vector $\underline{\mathbf{v}}_{0}$ = [$\mathbf{v}_{0}^{T}$ \ 0] is projected onto this hyperplane and $\underline{\mathbf{w}}_1$ is obtained as our first estimate as shown in Fig. \ref{app:convex1}. In the second step, $\underline{\mathbf{w}}_1$ is projected onto the set $\Csi$ by simply making its last component zero. The TV of this vector and the surface normal, and the supporting hyperplane are calculated as in the previous step. Next, $\underline{\mathbf{v}}_{0}$ is projected onto the new supporting hyperplane, and $\underline{\mathbf{w}}_{2}$ is obtained. In Fig. \ref{app:convex1}, $\underline{\mathbf{w}}_{2}$ is very close to the denoising solution $\underline{\mathbf{w}}^{\star}$. In general iterations continue until ${ \|\underline{\mathbf{w}}_{i} - \underline{\mathbf{w}}_{i-1}\|}\leq \epsilon$, where $\epsilon$ is a prescribed number, or iterations can be stopped after a certain number of iterations.

\begin{figure}[ht!]
\begin{center}
\noindent
\includegraphics[width=70mm]{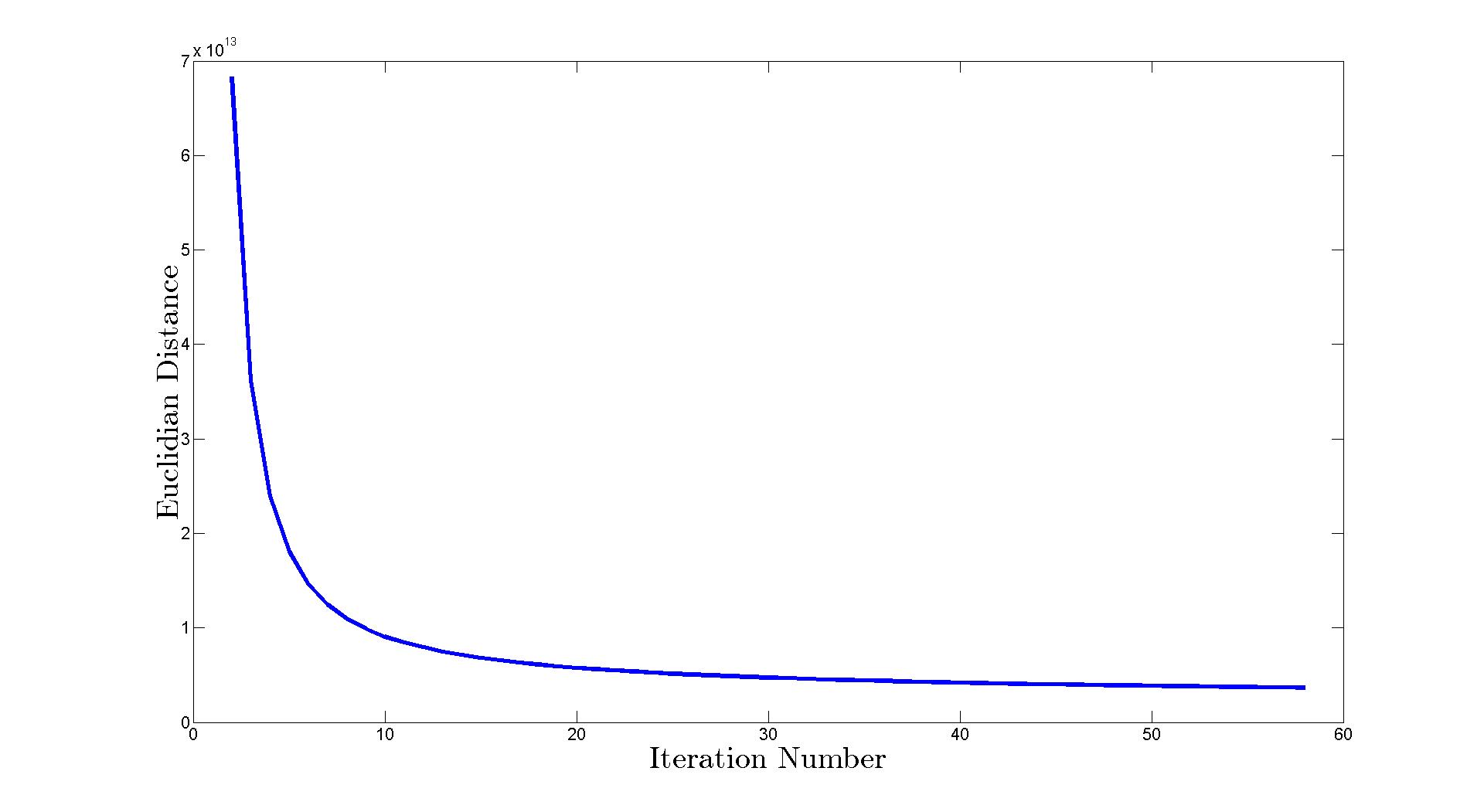}
\caption{Euclidian distance from \textbf{v} to the epigraph of TV at each iteration ($\|\underline{\mathbf{v}} - \underline{\mathbf{w}_i}\|$) with noise standard deviation of $\sigma=30$.}
\label{app:graphdist}
\end{center}
\end{figure}

We calculate the distance between $\underline{\mathbf{v}_{0}}$ and $\underline{\mathbf{w}}_i$ at each step of the iterative algorithm described in the previous paragraph. The distance ${ \|\underline{\mathbf{v}_{0}} - \underline{\mathbf{w}}_{i}\|}^2$ does not always decrease for high $i$ values. This happens around the optimal denoising solution $\underline{\mathbf{w}}^{\star}$. Once we detect an increase in ${ \|\underline{\mathbf{v}_{0}} - \underline{\mathbf{w}}_{i}\|}^2$, we perform a refinement step to obtain the final solution of the denoising problem. In refinement step, the supporting hyperplane at $\frac{\underline{\mathbf{w}}_{i}+\underline{\mathbf{w}}_{i-1}}{2}$ is used in the next iteration. A typical convergence graph is shown in Fig. \ref{app:graphdist} for the ``note" image.

It is possible to obtain a smoother version of $\underline{\mathbf{w}}^{\star}$ by simply projecting $\underline{\mathbf{v}}$ inside the set $\Cfi$ instead of the boundary of $\Cfi$.

\vspace{-0.1cm}
\section{Simulation Results}
\begin{figure}[ht!]
\begin{center}
\noindent
\includegraphics[width=80mm]{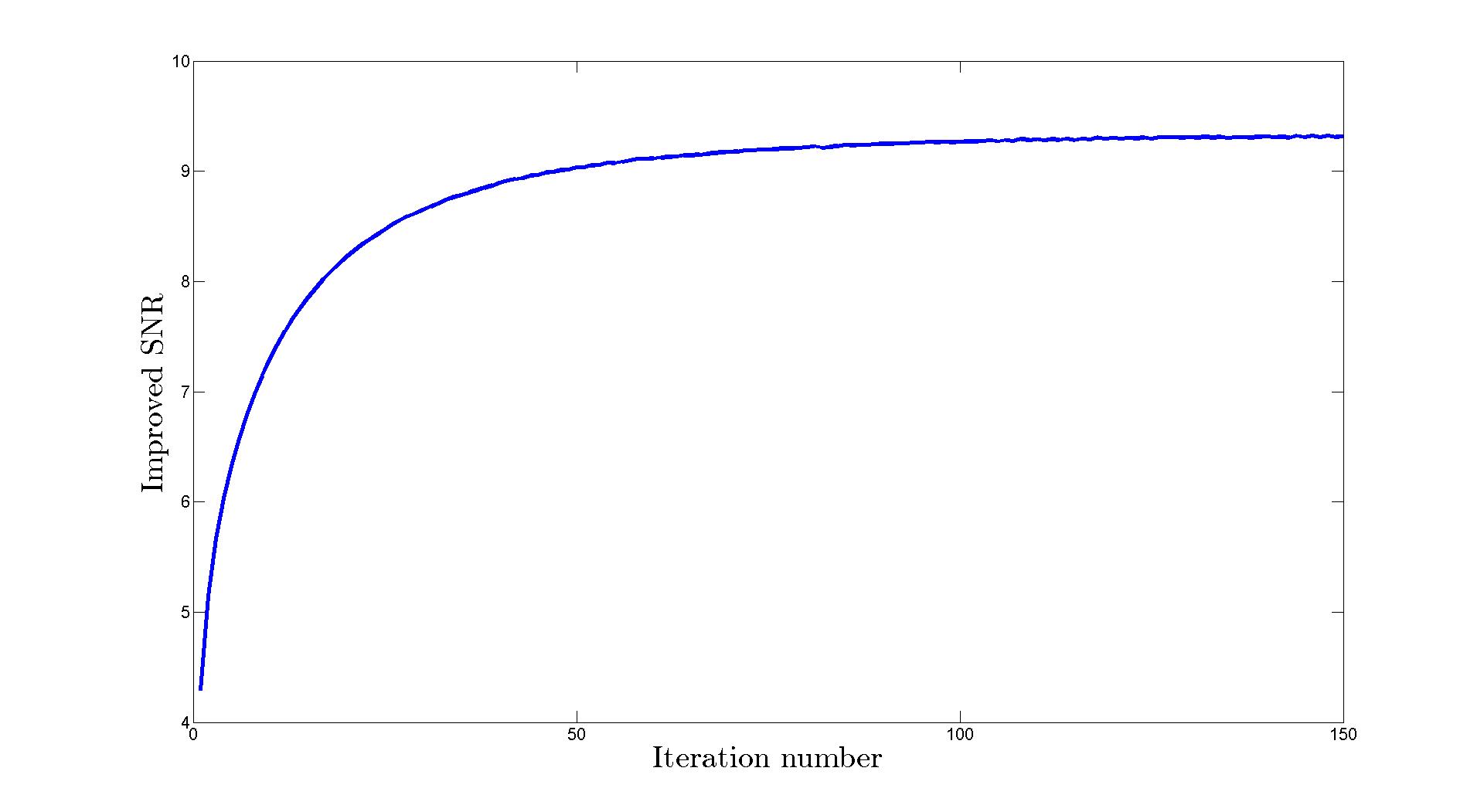}
\caption{ISNR as a function of the iteration number for MRI image.}
\label{app:graphISNR}
\end{center}
\end{figure}

\vspace{-0.1cm}
\label{sec:Simulation Results}
The PESC algorithm is tested with standard images. The noise standard deviation $\sigma$ is chosen so that the averaged blurred signal to noise ratio $BSNR$ reaches a target value:
\begin{equation}
\label{app:eq:eq10}
\textrm{BSNR} = 10\times log_{10}(\frac{\|\textbf{\~{z}}-E[\textbf{\~{z}}]\|^2}{N\sigma_\eta^2}),
\end{equation}
where $\textbf{\~{z}}$ is the blurred image without noise $\textbf{\~{z}} = \textbf{w}_{orig}\ast \textbf{h}$, $N$ is the whole number of pixels, and $\sigma$ is the additive noise's standard deviation. In addition to the visual results, the deblurring algorithm is compared in term of Improved Signal to Noise Ratio (ISNR) as follows:
\begin{equation}
\label{app:eq:eq11}
\textrm{ISNR} = 10\times log_{10}(\frac{\|\textbf{z}-\textbf{w}_{orig}\|^2}{\|\textbf{w}_{rec}-\textbf{w}_{orig}\|^2}),
\end{equation}
which $\textbf{w}_{rec}$ is the reconstructed and deblurred image. The ISNR as a function of iteration number for the experiment done over MRI image is given in Fig. \ref{app:graphISNR}.

Table \ref{tab:1} and \ref{tab:2} represent the ISNR and SNR values for five BSNR values for PESC algorithm and FTL algorithm proposed by Vonesch $et al$ \cite{Unser}. According to the these tables, in almost all cases PESC based deconvolution algorithm performs better than FTL \cite{Unser} in sense of ISNR and SNR.

\begin{table*}[ht!]
\begin{center}
\caption{ISNR and SNR results for PESC based deconvolution algorithm.}
\label{tab:1}
\scalebox{0.9} {
\begin{tabular}{|c|c|c|c|c|c|c|c|c|c|c|c|c|}
\hline
\parbox[t]{1cm}{BSNR }&\multicolumn{2}{|c|}{Cameraman}&\multicolumn{2}{|c|}{Lena}&\multicolumn{2}{|c|}{Peppers}&\multicolumn{2}{|c|}{Pirate}&\multicolumn{2}{|c|}{Mandrill}&\multicolumn{2}{|c|}{MRI}\\\hline\hline
30&5.59&20.83&4.48&24.94&5.35&26.13&4.57&22.77&4.56&20.77&4.64&13.71\\\hline
35&7.01&22.28&5.77&26.26&5.88&26.72&5.55&23.28&5.61&21.83&5.76&14.90\\\hline
40&8.49&23.77&6.95&27.46&7.45&28.32&6.75&24.99&6.41&22.65&7.07&16.28\\\hline
45&9.75&25.04&8.03&28.55&8.52&29.39&7.87&26.12&6.72&22.95&8.40&17.61\\\hline
50&10.76&26.10&8.49&29.00&9.50&30.37&8.41&26.66&6.84&23.07&9.31&18.53\\\hline
\end{tabular}}
\end{center}
\vspace{-0.3cm}
\end{table*}

\begin{table*}[ht!]
\begin{center}
\caption{ISNR and SNR results for FTL based deconvolution algorithm.}
\label{tab:2}
\scalebox{0.9} {
\begin{tabular}{|c|c|c|c|c|c|c|c|c|c|c|c|c|}
\hline
\parbox[t]{1cm}{BSNR }&\multicolumn{2}{|c|}{Cameraman}&\multicolumn{2}{|c|}{Lena}&\multicolumn{2}{|c|}{Peppers}&\multicolumn{2}{|c|}{Pirate}&\multicolumn{2}{|c|}{Mandrill}&\multicolumn{2}{|c|}{MRI}\\\hline\hline
30&-0.40&14.79&-0.44&19.70&-3.26&17.20&0.71&18.74&4.32&20.52&3.70&13.02\\\hline
35&6.16&21.35&5.78&25.97&5.66&26.11&5.61&23.68&5.45&21.65&5.05&14.36\\\hline
40&7.54&22.73&6.93&27.13&8.00&28.45&6.44&24.50&5.74&21.94&5.42&14.73\\\hline
45&7.89&23.08&7.26&27.46&8.56&29.02&6.67&24.75&5.89&22.08&5.55&14.86\\\hline
50&8.04&23.23&7.40&27.59&8.74&29.20&6.77&24.84&5.98&22.18&5.60&14.92\\\hline
\end{tabular}}
\end{center}
\vspace{-0.3cm}
\end{table*}

\begin{figure*}[ht]
\centering
\subfloat[Original]
{\label{fig:Original}\includegraphics[width=25mm]{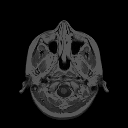}} \quad
\subfloat[Blurred]
{\label{fig:Blurred}\includegraphics[width=25mm]{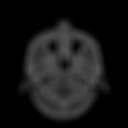}} \quad
\subfloat[PESC]
{\label{fig:PESC}\includegraphics[width=25mm]{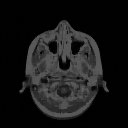}} \quad
\subfloat[FTL]
{\label{fig:FTL}\includegraphics[width=25mm]{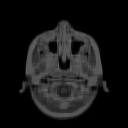}} \quad
\caption{Sample image used in our experiments (a) Original, (b) Blurred, (c) Deblurred by PESC, (d) Deblurred by FTL.}
\label{fig:otherOrg}
\end{figure*}

In Fig. \ref{fig:otherOrg} the original, blurred. and deblurred images for both algorithms are presented. According to this figure, PESC algorithm performs better than FTL not only in sense of SNR, but also the results for PESC are visually better than FTL.

\vspace{-0.2cm}
\section{Conclusion}
\vspace{-0.2cm}
\label{sec:Conclusion}
A new deconvolution method based on the epigraph of the TV function is developed. Epigraph sets of other convex cost functions can be also used in the new deconvolution approach. The reconstructed signal is obtained by making an orthogonal projection onto the epigraph set from the corrupted signal in $\mathbb{R}^{N+1}$. The new algorithm does not need the optimization of the regularization parameter as in standard TV deconvolution methods. Experimental results indicate that better SNR results are obtained compared to standard TV based deconvolution in a large range of images.
\begin{spacing}{0.9}
\setlength{\bibsep}{0.0pt}
\bibliographystyle{IEEEbib}
\bibliography{refs}

\begin{thebibliography}{10}

\bibitem{Bregman}
L.M. Bregman,
\newblock ``Finding the common point of convex sets by the method of successive
  projection.(russian),''
\newblock {\em \{USSR\} Doklady Akademii Nauk SSSR}, vol. 7, no. 3, pp. 200 --
  217, 1965.

\bibitem{You82}
D.C. Youla and H.~Webb,
\newblock ``Image {R}estoration by the {M}ethod of {C}onvex {P}rojections:
  {P}art 1 {N}um2014;theory,''
\newblock {\em IEEE Transactions on Medical Imaging}, vol. 1, pp. 81--94, 1982.

\bibitem{Bre67}
L.M. Bregman,
\newblock ``The {R}elaxation {M}ethod of {F}inding the {C}ommon {P}oint of
  {C}onvex {S}ets and {I}ts {A}pplication to the {S}olution of {P}roblems in
  {C}onvex {P}rogramming,''
\newblock {\em USSR Computational Mathematics and Mathematical Physics}, vol.
  7, pp. 200 -- 217, 1967.

\bibitem{Yin08}
W.~Yin, S.~Osher, D.~Goldfarb, and J.~Darbon,
\newblock ``Bregman iterative algorithms for $\ell_1$-minimization with
  applications to compressed sensing,''
\newblock {\em SIAM Journal on Imaging Sciences}, vol. 1, no. 1, pp. 143--168,
  2008.

\bibitem{Gunay}
O.~Gunay, K.~Kose, B.~U. Toreyin, and A.~E. Cetin,
\newblock ``Entropy-functional-based online adaptive decision fusion framework
  with application to wildfire detection in video,''
\newblock {\em IEEE Transactions on Image Processing}, vol. 21, pp. 2853--2865,
  2012.

\bibitem{GlobalSIP2013}
A.~E. Cetin, A.~Bozkurt, O.~Gunay, Y.~H. Habiboglu, K.~Kose, I.~Onaran, R.~A.
  Sevimli, and M.~Tofighi,
\newblock ``Projections onto convex sets (pocs) based optimization by
  lifting,''
\newblock {\em IEEE GlobalSIP}, 2013.

\bibitem{Yamada}
I.~Yamada S.~Ono, M.~Yamagishi,
\newblock ``A sparse system identification by using adaptively-weighted total
  variation via a primal-dual splitting approach,''
\newblock pp. 6029--6033, 2013.

\bibitem{Kose2013}
K.~Kose, O.~Gunay, and A.~E. Cetin,
\newblock ``Compressive sensing using the modified entropy functional,''
\newblock {\em Digital Signal Processing}, 2013.

\bibitem{Cen12}
Y.~Censor, W.~Chen, P.~L. Combettes, R.~Davidi, and G.~T. Herman,
\newblock ``On the {E}ffectiveness of {P}rojection {M}ethods for {C}onvex
  {F}easibility {P}roblems with {L}inear {I}nequality {C}onstraints,''
\newblock {\em Computational Optimization and Applications}, vol. 51, pp.
  1065--1088, 2012.

\bibitem{Sla08}
K.~Slavakis, S.~Theodoridis, and I.~Yamada,
\newblock ``Online {K}ernel-{B}ased {C}lassification {U}sing {A}daptive
  {P}rojection {A}lgorithms,''
\newblock {\em IEEE Transactions on Signal Processing}, vol. 56, pp.
  2781--2796, 2008.

\bibitem{Cetin89}
A.~E. Cetin,
\newblock ``Reconstruction of signals from fourier transform samples,''
\newblock {\em Signal Processing}, pp. 129--148, 1989.

\bibitem{Kose11}
K.~Kose and A.~E. Cetin,
\newblock ``Low-pass filtering of irregularly sampled signals using a set
  theoretic framework,''
\newblock {\em IEEE Signal Processing Magazine}, pp. 117--121, 2011.

\bibitem{Cen81}
Y.~Censor and A.~Lent,
\newblock ``An {I}terative {R}ow-{A}ction {M}ethod for {I}nterval {C}onvex
  {P}rogramming,''
\newblock {\em Journal of Optimization Theory and Applications}, vol. 34, pp.
  321--353, 1981.

\bibitem{Sla09}
S.~Konstantinos, S.~Theodoridis, and I.~Yamada,
\newblock ``Adaptive constrained learning in reproducing kernel hilbert spaces:
  the robust beamforming case,''
\newblock {\em IEEE Transactions on Signal Processing}, vol. 57, pp.
  4744--4764, 2009.

\bibitem{The11}
K.~S Theodoridis and I.~Yamada,
\newblock ``Adaptive learning in a world of projections,''
\newblock {\em IEEE Signal Processing Magazine}, vol. 28, no. 1, pp. 97--123,
  2011.

\bibitem{censor1987optimization}
Y.~Censor and A.~Lent,
\newblock ``Optimization of �$\backslash$logx� entropy over linear equality
  constraints,''
\newblock {\em SIAM Journal on Control and Optimization}, vol. 25, no. 4, pp.
  921--933, 1987.

\bibitem{Tru85}
H.~J. Trussell and M.~R. Civanlar,
\newblock ``The {L}andweber {I}teration and {P}rojection {O}nto {C}onvex
  {S}et,''
\newblock {\em IEEE Transactions on Acoustics, Speech and Signal Processing},
  vol. 33, no. 6, pp. 1632--1634, 1985.

\bibitem{Com04}
P.~L. Combettes and J.-Ch. Pesquet,
\newblock ``Image restoration subject to a total variation constraint,''
\newblock {\em IEEE Transactions on Image Processing}, vol. 13, pp. 1213--1222,
  2004.

\bibitem{Com93}
P.~L. Combettes,
\newblock ``The foundations of set theoretic estimation,''
\newblock {\em Proceedings of the IEEE}, vol. 81, pp. 182 --208, 1993.

\bibitem{yamada2011minimizing}
I.~Yamada, M.~Yukawa, and M.~Yamagishi,
\newblock ``Minimizing the moreau envelope of nonsmooth convex functions over
  the fixed point set of certain quasi-nonexpansive mappings,''
\newblock {\em Springer NY}, pp. 345--390, 2011.

\bibitem{censor1987some}
Y.~Censor and G.~T. Herman,
\newblock ``On some optimization techniques in image reconstruction from
  projections,''
\newblock {\em Applied Numerical Mathematics}, vol. 3, no. 5, pp. 365--391,
  1987.

\bibitem{Sez82}
I.~Sezan and H.~Stark,
\newblock ``Image restoration by the method of convex projections: Part
  2-applications and numerical results,''
\newblock {\em IEEE Transactions on Medical Imaging}, vol. 1, pp. 95--101,
  1982.

\bibitem{censor1992proximal}
Y.~Censor and S.~A. Zenios,
\newblock ``Proximal minimization algorithm withd-functions,''
\newblock {\em Journal of Optimization Theory and Applications}, vol. 73, pp.
  451--464, 1992.

\bibitem{Tuy81}
A.~Lent and H.~Tuy,
\newblock ``An {I}terative {M}ethod for the {E}xtrapolation of {B}and-{L}imited
  {F}unctions,''
\newblock {\em Journal of Optimization Theory and Applications}, vol. 83, pp.
  554--565, 1981.

\bibitem{censor1981row}
Y.~Censor,
\newblock ``Row-action methods for huge and sparse systems and their
  applications,''
\newblock {\em SIAM review}, vol. 23, pp. 444--466, 1981.

\bibitem{censor1991optimization}
Y.~Censor, A.~R De~Pierro, and A.~N. Iusem,
\newblock ``Optimization of burg's entropy over linear constraints,''
\newblock {\em Applied Numerical Mathematics}, vol. 7, no. 2, pp. 151--165,
  1991.

\bibitem{Ros13}
M.~Rossi, A.~M. Haimovich, and Y.~C. Eldar,
\newblock ``Conditions for {T}arget {R}ecovery in {S}patial {C}ompressive
  {S}ensing for {MIMO} {R}adar,''
\newblock {\em IEEE ICASSP}, 2013.

\bibitem{Gub67}
L.G. Gubin, B.T. Polyak, and E.V. Raik,
\newblock ``The {M}ethod of {P}rojections for {F}inding the {C}ommon {P}oint of
  {C}onvex {S}ets,''
\newblock {\em Computational Mathematics and Mathematical Physics}, vol. 7, pp.
  1 -- 24, 1967.

\bibitem{Cet97}
A.~E. \c{C}etin, O.N. Gerek, and Y.~Yardimci,
\newblock ``Equiripple {FIR} {F}ilter {D}esign by the {FFT} {A}lgorithm,''
\newblock {\em IEEE Signal Processing Magazine}, vol. 14, no. 2, pp. 60--64,
  1997.

\bibitem{Chambolle}
A.~Chambolle,
\newblock ``An algorithm for total variation minimization and applications,''
\newblock {\em Journal of Mathematical Imaging and Vision}, vol. 20, no. 1-2,
  pp. 89--97, Jan. 2004.

\bibitem{Kos12}
K.~Kose, V.~Cevher, and A.E. Cetin,
\newblock ``Filtered variation method for denoising and sparse signal
  processing,''
\newblock {\em IEEE ICASSP}, pp. 3329--3332, 2012.

\bibitem{Bar07}
R.G. Baraniuk,
\newblock ``Compressive sensing [lecture notes],''
\newblock {\em IEEE Signal Processing Magazine}, vol. 24, pp. 118--121, 2007.

\bibitem{Com11}
P.~L. Combettes and J.-Ch. Pesquet,
\newblock ``Proximal splitting methods in signal processing,''
\newblock Springer Optimization and Its Applications, pp. 185--212. Springer
  NY, 2011.

\bibitem{Com123}
P.L. Combettes and J.~Pesquet,
\newblock ``Image deconvolution with total variation bounds,''
\newblock in {\em Signal Processing and Its Applications, 2003. Proceedings.
  Seventh International Symposium on}, July 2003, vol.~1, pp. 441--444 vol.1.

\bibitem{Pesquet12}
Giovanni Chierchia, Nelly Pustelnik, Jean-Christophe Pesquet, and B{\'e}atrice
  Pesquet-Popescu,
\newblock ``Epigraphical projection and proximal tools for solving constrained
  convex optimization problems: Part i,''
\newblock {\em CoRR}, vol. abs/1210.5844, 2012.

\bibitem{Unser}
C.~Vonesch and M.~Unser,
\newblock ``A fast thresholded landweber algorithm for wavelet-regularized
  multidimensional deconvolution,''
\newblock {\em Image Processing, IEEE Transactions on}, vol. 17, no. 4, pp.
  539--549, April 2008.

\end{thebibliography}
\end{spacing}

\end{document}